\documentclass[12pt]{article} 

\usepackage{ol2}
\usepackage[draft]{hyperref}
\usepackage{amsmath}

\begin{document}



\title{Photon-number correlations by photon-number resolving detectors}
\author{Alessia Allevi,$^1$ Maria Bondani,$^{2,*}$ and Alessandra Andreoni$^{3}$}
\address{$^1$ C.N.I.S.M., U.d.R. Milano, I-20133, Italy}
\address{$^2$ National Laboratory for Ultrafast and Ultraintense
Optical Science - C.N.R.-I.N.F.M., Como, I-22100, Italy}
\address{$^3$ Dipartimento di Fisica e Matematica, Universit\`a degli Studi dell'Insubria and C.N.I.S.M., U.d.R. Como, I-22100, Italy}
\address{$^*$Corresponding author: maria.bondani@uninsubria.it}
\begin{abstract}
We demonstrate that by using a pair of photodetectors endowed with internal gain we are
able to quantify the correlation coefficient between the two components of a pulsed
bipartite state in the mesoscopic intensity regime (less than 100 mean photons).
\end{abstract}

\begin{figure}[htbp]
  \centering
  \includegraphics[angle=0,width=1\textwidth]{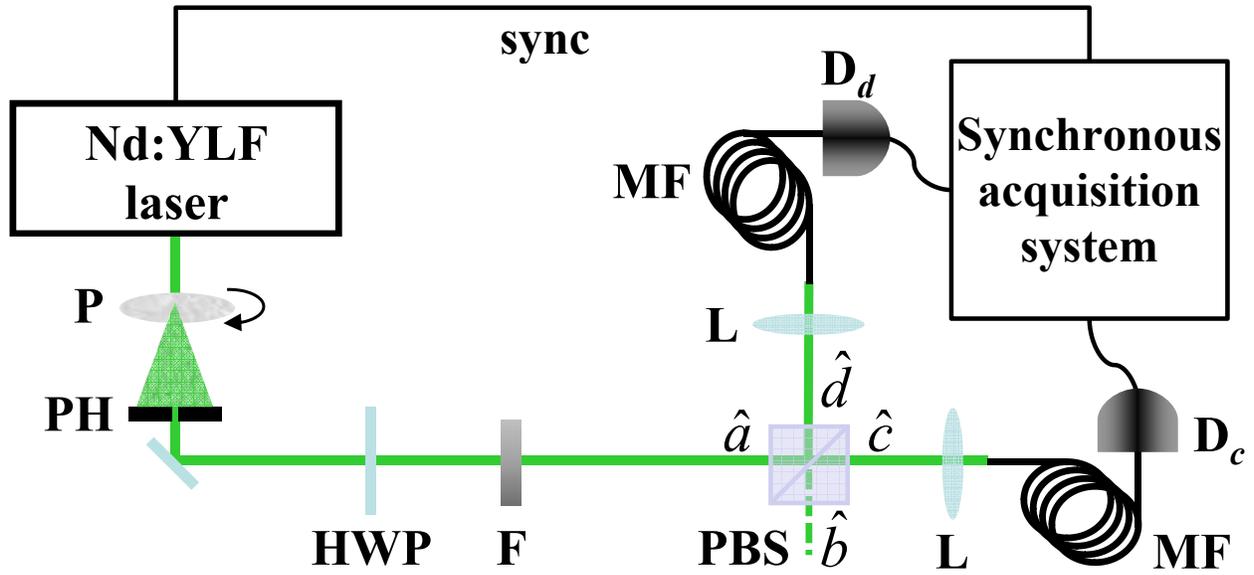}
  \caption{(Color online) Scheme of the experimental setup. P, rotating diffuser plate; PH, pin-hole;
  F, variable neutral density filter; HWP, half wave plate;
  PBS, polarizing beam splitter; L, lenses; MF, multimode optical fibers;
  D$_c$ and D$_d$, photo-detectors. }\label{f:setup}
  \end{figure}

\begin{figure}[htbp]
  \centering
  \includegraphics[angle=0,width=1\textwidth]{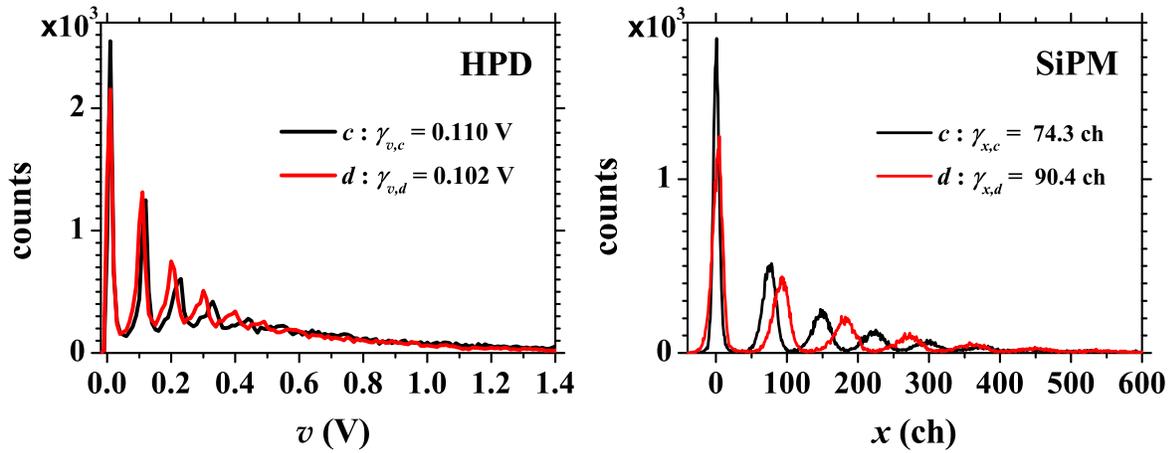}
  \caption{(Color online) Pulse height spectra at the outputs of the beam splitter as
  measured by HPD and SiPM photodetectors at very similar intensities. }\label{f:spectra}
  \end{figure}

\begin{figure}[htbp]
  \centering
  \includegraphics[angle=0,width=1\textwidth]{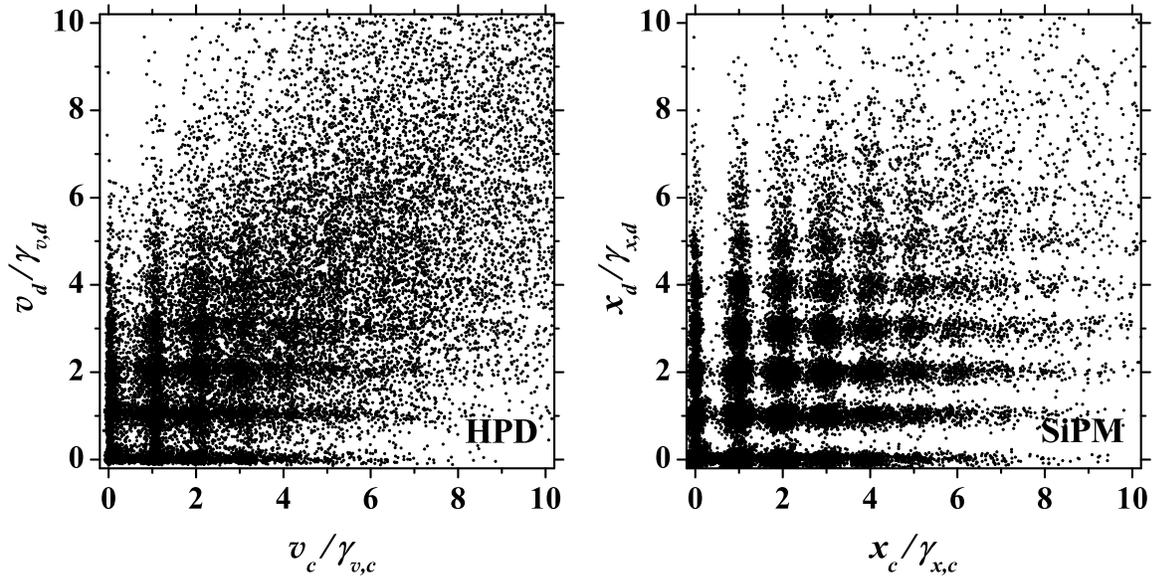}
  \caption{(Color online) Experimental joint probability distribution. }\label{f:conj}
  \end{figure}

\begin{figure}[htbp]
  \centering
  \includegraphics[angle=0,width=1\textwidth]{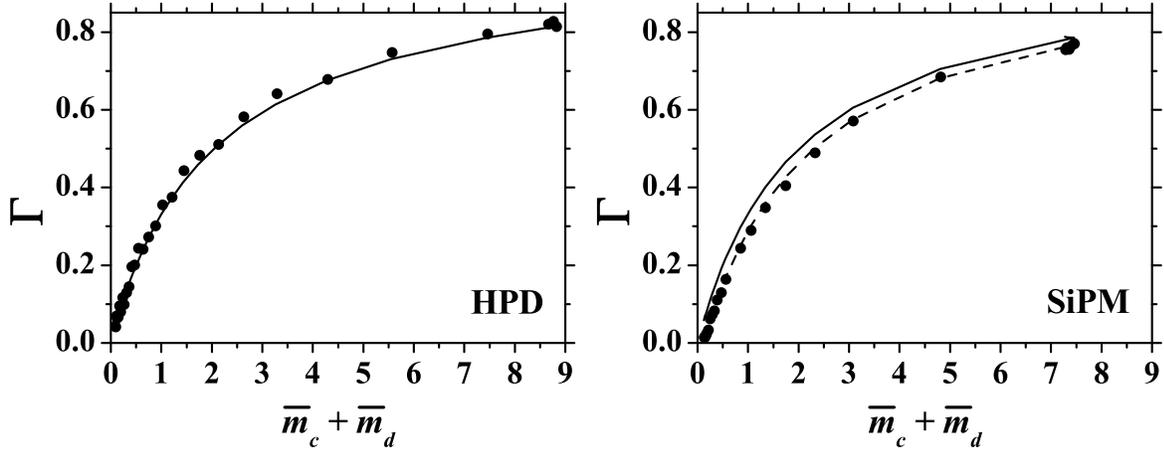}
  \caption{(Color online) Shot-by-shot correlation, $\Gamma$, as a function of the mean value
  of output detected light (dots) together with the theoretical curves (full lines).
  The dashed line in the right panel is the theoretical prediction corrected for dark counts
  ($\overline{m}_{dc,c}= 0.071$, $\overline{m}_{dc, d}= 0.034$) and cross talk
  ($\epsilon_c = 0.10$, $\epsilon_d = 0.08$).}\label{f:corr}
  \end{figure}

\begin{figure}[htbp]
  \centering
  \includegraphics[angle=0,width=1\textwidth]{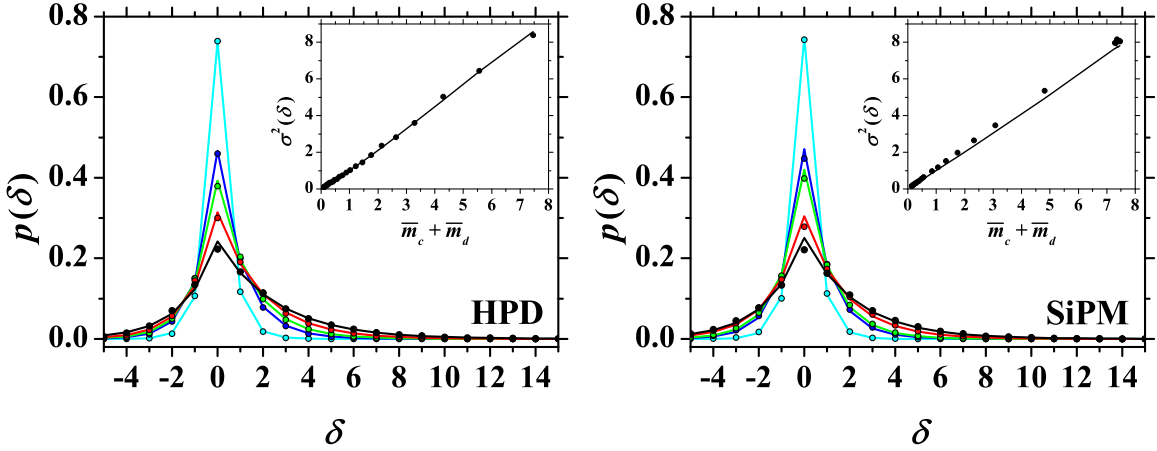}
  \caption{(Color online) Experimental results for $p(\delta)$ (full circles) and theoretical
  predictions (full lines). The fidelities of the data are
  $f=0.9999,0.9997,0.9996,0.9994,0.9991$, for HPDs and $f=0.9997,0.9994,0.9994,0.9993,0.9991$, for SiPM.
  Insets: measured variance $\sigma^2(\delta)$ of $p(\delta)$ as a function of the
  total number of detected photons (dots) and theoretical predictions (full line). }\label{f:differ}
  \end{figure}

The existence of photon-number correlations is necessary to produce conditional states
based on number photodetection. Moreover the study of the properties of photon-number
correlations and, in particular, of the distribution of the difference in the number of
photons detected in the two components of a pulsed bipartite state, is meaningful of the
properties of the state \cite{jointdiff,ivo}. For instance, measurements of sub-shot
noise correlations have been used to determine the nonclassicality of both bipartite
\cite{PRAsubshot,ottavia,chekhovaARCH} and tripartite states \cite{PRAtribit}.
\\
Different types of commercial detectors can be used to measure photon numbers in the
different intensity regimes: pin photodiodes \cite{kumar,jointdiff}, amplified pin
photodiodes \cite{PRAsubshot,PRAtribit,chekhovaARCH}, CCD's
\cite{ottavia,haderka,torino}, loop detectors \cite{micuda,silberhorn}, hybrid
photodetectors \cite{selfconsistent,ASL,OLwigner}, multipixel photodetectors
\cite{silberberg,ramilli}. Some of them have been exploited to reconstruct photon-number
statistics
\cite{haderka,PRAsubshot,PRAtribit,silberhorn,selfconsistent,ASL,OLwigner,silberberg,ramilli}
or for correlation measurements
\cite{haderka,chekhovaARCH,PRAsubshot,PRAtribit,ottavia,torino}.
\\
We have already demonstrated that hybrid photodiode modules (HPD, R10467U-40, Hamamatsu,
maximum quantum efficiency $\sim0.4$ at 550~nm) and Silicon photomultipliers (SiPM, MPPC
S10362-11-100C, Hamamatsu, maximum quantum efficiency $\sim 0.65$ at 440~nm) are endowed
with the capability of reconstructing the correct statistics of detected photons provided
their outputs are analyzed in the proper way \cite{selfconsistent,ASL,OLwigner,ramilli}.
In this Letter we exploit the fact that, in principle, our analysis allows us to
determine the number of photons detected at each laser shot and demonstrate that pairs of
the above detectors can be used to measure shot-by-shot photon-number correlations
between the components of a bipartite state. To test the detection procedure, we start by
measuring the correlations in a classically-correlated bipartite state, namely a
single-mode thermal field divided by a beam-splitter.
%
%
%
\\
The experimental setup is sketched in Fig.~\ref{f:setup}. The single-mode thermal field
is obtained by passing the second-harmonic pulses ($\sim$5.4~ps, 523~nm) of a Nd:YLF
mode-locked laser amplified at 500 Hz (High Q Laser Production) through a rotating
ground-glass plate (P in Fig.~\ref{f:setup}) \cite{Arecchi1965} and selecting a single
coherence area in the far field by a pin-hole (PH, 150~$\mu$m diameter). The selected
mode is then sent to a polarizing beam-splitter (PBS) whose outputs are collected by two
multimode fibers and delivered to the detectors. To obtain a fine tuning of the
beam-splitter transmittance we inserted a half-wave plate (HWP) between the PH and the
PBS. Single-shot detector outputs were then amplified, integrated, digitized and
recorded. We measured 30000 subsequent shots with the HPD detectors and 50000 with the
SiPM detectors at a number of different intensity values set by a variable filter (F).
Figure~\ref{f:spectra} shows two typical pulse height spectra obtained at similar
intensities: clearly, the response of the SiPMs produces better resolved peaks as
compared to the HPDs. As the HPDs are not affected by dark-counts, by applying the
selfconsistent procedure described in \cite{selfconsistent} to the output of the
detectors, we can evaluate the number of detected photons in each laser shot. The
procedure gives the value of the overall gain ($\gamma$) of the detection apparatus that
links the final outputs (voltages for HPDs and multichannel addresses for SiPM) to the
number of detected photons. In the case of SiPMs, the procedure described in
\cite{ramilli} yields a value of detected photons that includes the unavoidable effects
of dark-counts and cross-talk, that must be taken into account to properly interpret the
experimental results. The procedure itself yields the values of the dark-count rate
$\overline{m}_{dc}$ and of the cross-talk probability $\epsilon$ for the considered
data-set.
%
%
%
\\
In Fig.~\ref{f:conj} we plot the joint probability distributions obtained by re-scaling
the outputs of the detectors by the proper value of $\gamma$ but without rebinning the
values. As expected, the results for SiPM display a better localization of the peaks,
which should promise a better determination of the correlation.
%
%
%
\\
Referring to Fig.~\ref{f:setup}, the field operators at the outputs of the beam splitter
($\hat{c}$ and $\hat{d}$) are linked to the input ones ($\hat{a}$ and $\hat{b}$) by:
\begin{equation}
   \label{eq:bs}
   \hat{c} =  \sqrt{\tau} \, \hat{a} - \sqrt{1-\tau} \, \hat{b}\;\;,\;\;
   \hat{d} =  \sqrt{1-\tau} \, \hat{a} + \sqrt{\tau}\,\hat{b}\; .
\end{equation}
If $\hat{b}$ is in the vacuum, the moments of the bipartite state at the output are given
by $\langle\hat{n}_c\rangle = \tau \langle\hat{n}_a\rangle$, $\sigma^2_{n_c}=\tau
\left[\tau\sigma^2_{n_a} + (1-\tau) \langle\hat{n}_a\rangle\right]$,
$\langle\hat{n}_d\rangle= (1-\tau) \langle\hat{n}_a\rangle$, $\sigma^2_{n_d}= (1-\tau)
\left[(1-\tau)\sigma^2_{n_a} + \tau \langle\hat{n}_a\rangle\right]$ and
$\langle\hat{n}_c\,\hat{n}_d\rangle= \tau(1-\tau) \left(\langle\hat{n}^2_a\rangle -
\langle\hat{n}_a\rangle\right)$, from which we obtain the expression for the normalized
correlation coefficient:
\begin{eqnarray}
   \label{eq:gamma}
   \Gamma &\equiv&\frac{\langle\left({n}_c-\langle\hat{n}_c\rangle\right)\left({n}_d-
   \langle\hat{n}_d\rangle\right)\rangle}{\sigma^2_{n_c}\sigma^2_{n_d}}\nonumber\\
   &=& \frac{\sqrt{\tau(1-\tau)} \left[\sigma^2_{n_a} - \langle{\hat{n}_a}\rangle\right] }
   {\sqrt{\left[\tau\sigma^2_{n_a} + (1-\tau)\langle{\hat{n}_a}\rangle\right]
   \left[(1-\tau)\, \sigma^2_{n_a} + \tau \langle{\hat{n}_a}\rangle\right]}}\; .
\end{eqnarray}
\\
For a single-mode thermal state at the input, $\hat{\rho}_a = \sum_n p_{\mathrm{th}}(n)
|n\rangle{n}\langle n|$ with $p_{\mathrm{th}}(n) = \overline{n}^n/(1+\overline{n})^{1+n}$
and $\langle\hat{n}_a\rangle=\overline{n}$, we have
\begin{equation}
   \label{eq:gammaT}
   \Gamma = \frac{\sqrt{\tau(1-\tau)\overline{n}^2}}{\sqrt{(\tau\overline{n}+1)[(1-\tau)\overline{n}+1]}}
   =\frac{\sqrt{\overline{n}_c\overline{n}_d}}{\sqrt{(\overline{n}_c+1)(\overline{n}_d+1)}},
\end{equation}
where $\overline{n}_c=\langle\hat{n}_c\rangle$ and
$\overline{n}_d=\langle\hat{n}_d\rangle$. If the primary detection event is performed
with quantum efficiency $\eta$ and no dark counts, the expression in
Eq.~(\ref{eq:gammaT}) remains the same upon substituting the mean number of photons
($\overline{n}_{c,d}$) with that of detected photons
($\overline{m}_{c,d}=\eta_{c,d}\overline{n}_{c,d}$)\cite{jointdiff}.
\\
Figure~\ref{f:corr} shows the results for shot-by-shot correlation measured with HPDs and
SiPM (dots) as a function of the mean value of output detected light
($\overline{m}_c+\overline{m}_d$). The full lines are the theoretical values evaluated
according to Eq.~(\ref{eq:gammaT}) by inserting the experimental mean values. Not
surprisingly the results obtained with HPDs perfectly superimpose to the theory, while
those obtained with SiPMs display a reduced correlation, due to the effect of dark-counts
and cross talk. By evaluating the theoretical curve at the effective mean values
$\overline{m}_{eff}=(\overline{m}-\overline{m}_{dc})/(1+\epsilon)$, we obtain the dashed
curve that fits the experimental data rather satisfactorily.
%
%
\\
Relevant applications of correlated bipartite states require the determination of the
nature of the correlations. A quite standard way to discriminate between classical and
nonclassical states \cite{jointdiff} is to study the statistics of the difference,
$p(\delta)$, in the number of photons detected shot-by-shot at the outputs of the beam
splitter. Figure~\ref{f:differ} displays the experimental results for $p(\delta)$
evaluated by subtracting the number of detected photons measured shot-by-shot with HPDs
and SiPMs together with the theoretical values evaluated according to Ref.~[1]. The
sample curves where chosen to have very similar detected mean values for the two classes
of detectors. By evaluating the fidelity
($f=\sum_{i}\sqrt{p_i(\delta)p_i^{\mathrm{th}}(\delta)}$) we note that quality of the
results is almost the same for HPDs and SiPMs, although for the first is slightly better
(see figure caption for the values of $f$). The Insets of Figure~\ref{f:differ} display
the measured variance $\sigma^2(\delta)$ as a function of the total number of detected
photons, as compared to the theoretical predictions $\sigma^2(\delta) = (\overline{m}_c -
\overline{m}_d)^2 +\overline{m}_c+\overline{m}_d$ \cite{jointdiff}. The correct
evaluation of $\sigma^2(\delta)$ is crucial for the estimation of the noise reduction in
a bipartite state $R=\sigma^2(\delta)/(\overline{m}_c+\overline{m}_d)$, which is a means
to discriminate classical from nonclassical states. Again the results obtained with HPDs
superimpose to theory, whereas those obtained with SiPM follow a broader distribution:
these results indicate that HPDs can give direct information on sub-shot noise level,
whereas the information obtained with SiPM must be corrected to be interpreted.
%
%
\\
In conclusion we have demonstrated that detectors partially endowed with photon-number
resolution can be used to measure photon-number correlations shot-by-shot and to reliably
estimate the noise reduction factor. We have also demonstrated that, although endowed
with a better photon-number resolution, SiPM are affected by spurious effects
(dark-counts and cross-talk) that degrade the quality of results and require corrections
that risk to prevent their use for state preparation by conditional measurements
\cite{silberberg}. However, with both kinds of detectors our system is less cumbersome
than that described in Ref.~[10] and, in the case of HPD, provides similar overall
detection efficiency.


The Authors thank M. Caccia and his group (Insubria University) for the loan of the SiPM
and of some electronic equipment, F. Beduini (ICFO Barcelona) and M. Ramilli (Insubria
University) for experimental assistance, and M. G. A. Paris (Milano University) for
fruitful discussions.
\\
Present address of Alessia Allevi is C.N.I.S.M., U.d.R. Como, I-22100, Italy.


\clearpage

\section*{List of Figure Captions}

\noindent Fig. 1. (Color online) Scheme of the experimental setup. P, rotating diffuser
plate; PH, pin-hole;  F, variable neutral density filter; HWP, half wave plate;  PBS,
polarizing beam splitter; L, lenses; MF, multimode optical fibers; D$_c$ and D$_d$,
photo-detectors.

\noindent Fig. 2. (Color online) Pulse height spectra at the outputs of the beam splitter
as measured by HPD and SiPM photodetectors at very similar intensities. %

\noindent Fig. 3. (Color online) Experimental joint probability distribution.
\noindent Fig. 4. (Color online) Shot-by-shot correlation, $\Gamma$, as a function of the
mean value of output detected light (dots) together with the theoretical curves (full
lines). The dashed line in the right panel is the theoretical prediction corrected for
dark counts ($\overline{m}_{dc,c}= 0.071$, $\overline{m}_{dc, d}= 0.034$) and cross talk
($\epsilon_c = 0.10$, $\epsilon_d = 0.08$).
\noindent Fig. 5. (Color online) Experimental results for $p(\delta)$ (full circles) and
theoretical predictions (full lines). The fidelities of the data are
$f=0.9999,0.9997,0.9996,0.9994,0.9991$, for HPDs and
$f=0.9997,0.9994,0.9994,0.9993,0.9991$, for SiPM. Insets: measured variance
$\sigma^2(\delta)$ of $p(\delta)$ as a function of the total number of detected photons
(dots) and theoretical predictions (full line).
\clearpage
\clearpage

\clearpage

\clearpage

\clearpage

\end{document}